\def\year{2018}\relax
\documentclass[letterpaper]{article} 
\usepackage{aaai18}  
\usepackage{times}  
\usepackage{helvet}  
\usepackage{courier}  
\usepackage{url}  
\usepackage{graphicx}  
\usepackage[dvipsnames]{xcolor}
\usepackage{color,xcolor,natbib,rotating}
\let\cite=\citep
\usepackage{silence}
\usepackage{subcaption}
\usepackage{enumitem}

\newcommand{\ra}[1]{\renewcommand{\arraystretch}{#1}}
\usepackage{booktabs,multirow,tabularx}
\usepackage{makecell}

\newcommand{\xhdr}[1]{\vspace{1mm}\noindent{{\bf #1.}}}

\usepackage{etoolbox}
\makeatletter
\patchcmd{\maketitle}{\copyright@on}{}{}{}
\makeatother

\graphicspath{{./figures/}}

\begin{document}

\title{Do Diffusion Protocols Govern Cascade Growth?}
\author{Justin Cheng$^{1}$, Jon Kleinberg$^{2}$, Jure Leskovec$^{3}$, David Liben-Nowell$^{4}$,\\
{\bf \Large Bogdan State$^{1}$, Karthik Subbian$^{1}$, and Lada Adamic$^{1}$}\\
jcheng@fb.com, kleinber@cs.cornell.edu, jure@cs.stanford.edu, dln@carleton.edu,\\
bogdanstate@fb.com, ksubbian@fb.com, ladamic@fb.com\\
$^{1}$Facebook, $^{2}$Cornell University, $^{3}$Stanford University, $^{4}$Carleton College
}
\maketitle

\begin{abstract}

Large cascades can develop in online social networks as people share information with one another.
Though simple reshare cascades have been studied extensively, the full range of cascading behaviors on social media is much more diverse.
Here we study how {\em diffusion protocols}, or the social exchanges that enable information transmission, affect cascade growth, analogous to the way communication protocols define how information is transmitted from one point to another.
Studying 98 of the largest information cascades on Facebook, we find a wide range of diffusion protocols -- from cascading reshares of images, which use a simple protocol of tapping a single button for propagation, to the ALS Ice Bucket Challenge, whose diffusion protocol involved individuals creating and posting a video, and then nominating specific others to do the same.
We find recurring classes of diffusion protocols, and identify two key counterbalancing factors in the construction of these protocols, with implications for a cascade's growth: the effort required to participate in the cascade, and the social cost of staying on the sidelines.
Protocols requiring greater individual effort slow down a cascade's propagation, while those imposing a greater social cost of not participating increase the cascade's adoption likelihood.
The predictability of transmission also varies with protocol.
But regardless of mechanism, the cascades in our analysis all have a similar reproduction number ($\approx$1.8), meaning that lower rates of exposure can be offset with higher per-exposure rates of adoption.
Last, we show how a cascade's structure can not only differentiate these protocols, but also be modeled through branching processes.
Together, these findings provide a framework for understanding how a wide variety of information cascades can achieve substantial adoption across a network.

\end{abstract}

\section{Introduction}

\begin{figure}[t]
  \centering
  \includegraphics[width=0.75\columnwidth,clip]{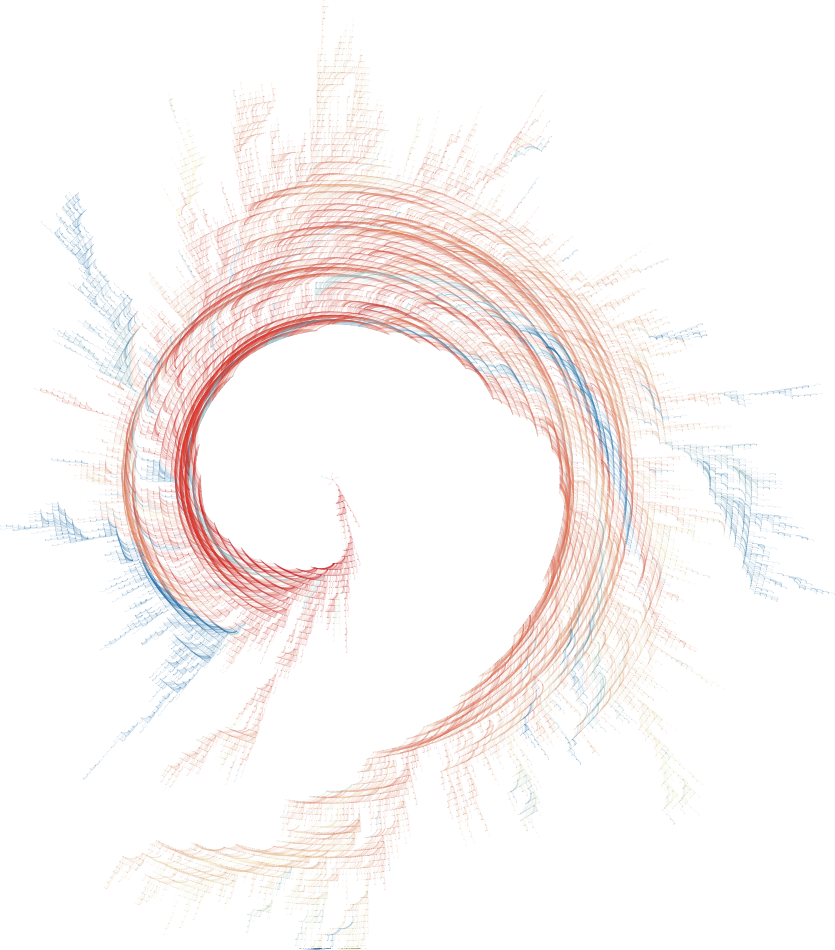}
  \caption{The diffusion tree of a cascade with a volunteer diffusion protocol, where individuals posted music from an artist whose name matched the letter they were assigned by a friend. Edges are colored from red (early) to blue (late).}
  \label{fig:music_challenge}
\end{figure}

On social network platforms such as Facebook, people reshare information they see from both friends and pages (i.e., the accounts of brands, organizations, or public figures).  This resharing may prompt other friends or pages to share as well, leading to a cascade that spreads through the network.
For these information cascades to grow large, the information within them must continually be replicated.
Thus, one may expect that large cascades should be those whose information is easily transmitted (e.g., reshareable with the tap of a button) and which spread through highly connected, central nodes in a network (e.g., pages with millions of followers) \cite{goldenberg2009role, hinz2011seeding}.

\begin{figure*}[t]
  \centering
  \includegraphics[width=\textwidth]{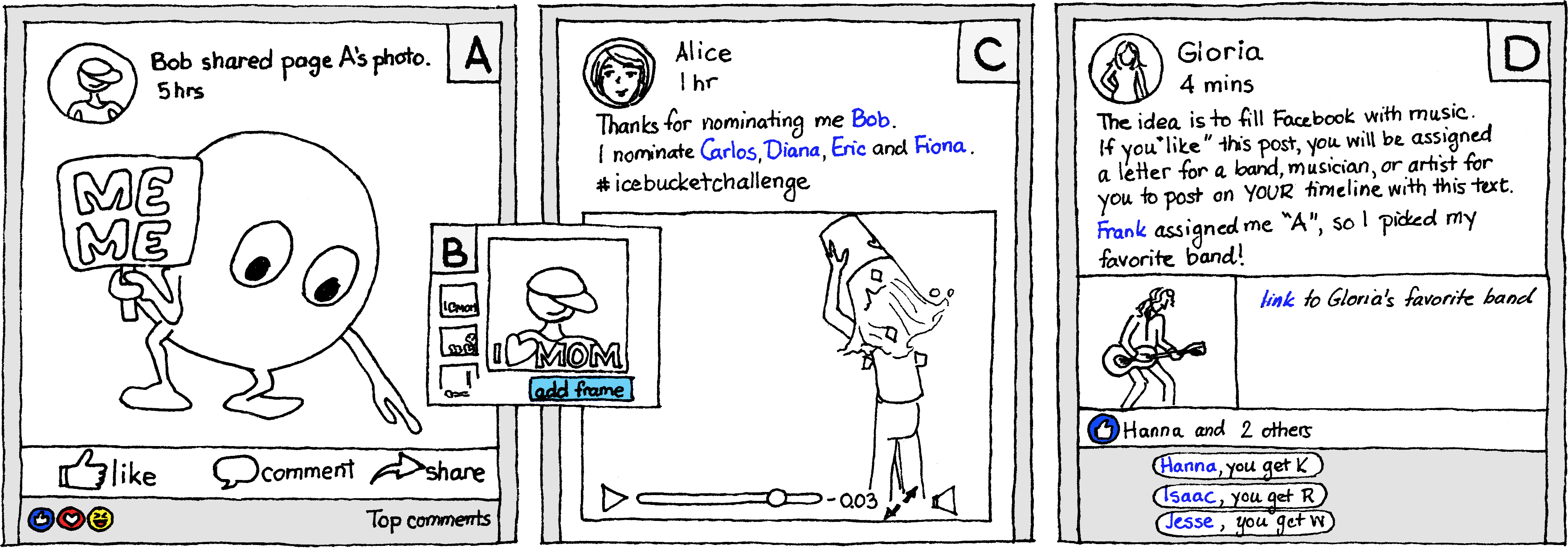}
  \caption{Four primary diffusion protocols in cascading behavior on Facebook. (Names are fictitious.)  (a, b) Transient and persistent copy protocols involve simple content replication. (c) Nomination protocols require participants to complete a task before inviting specific others to participate. (d) Volunteer protocols invite others to declare a desire to participate in an activity.}
  \label{fig:examples}
\end{figure*}

Unsurprisingly, the most common large cascades on Facebook consist of photo, video, or link reshares.
But alongside these simpler cascades, information and behavior is also spreading through more complex and effortful propagation mechanisms.
These more complex cascades are less frequent by comparison, but can still grow large.
The ALS Ice Bucket Challenge, for example, required significant effort to spread from one person to another: one participates by posting a video of oneself pouring ice water over one's head, and typically participates only if one was explicitly nominated by an earlier participant.
Nonetheless, the Ice Bucket Challenge resulted in $\mathop{>}$17M videos shared \cite{fbicebucketchallenge}.
As another example, a music challenge meme (Fig.~\ref{fig:music_challenge}, \ref{fig:examples}d) required people to volunteer for and then be assigned to post about a specific music video, with the resulting cascade attracting $\mathop{>}$200,000 participants.
Why did these cascades succeed despite requiring more effort and being limited in membership?
By studying the diverse social mechanisms by which large cascades propagate, we can gain a better understanding of the factors that govern cascade growth.

\xhdr{The present work: Diffusion protocols}
Here we study \emph{diffusion protocols,} the social exchanges that enable information to be transmitted in cascades, and how such protocols explain the substantial diversity in cascade structure and growth.
Similar to the way that different communication protocols define how information is transmitted between two endpoints in a computer network \cite{comer2000internetworking}, diffusion protocols define the interactions needed for information to spread across edges in a social network.

By collecting a variety of large cascades over multiple periods, we identify four types of underlying mechanisms (Fig.~\ref{fig:examples}).
These protocols can be characterized using two counterbalancing dimensions:  the \emph{individual effort} required to participate in a cascade, and the \emph{social cost} of not participating.
Each class of protocol induces a different combination of effort and social cost of not participating:
\begin{itemize}
\item A {\em transient copy protocol} underlies reshare and copy-paste cascades, where an individual simply copies the information, e.g., by tapping on a ``reshare'' button.
Here, the individual effort to participating is low and there is effectively zero social cost to not participating.
\item A {\em persistent copy protocol} differs from the first in its effect being persistent -- that is, visible for a longer time -- rather than transient like a post that quickly goes by.
Examples include cascading modifications to users' profile pictures; for example, adopting an item such as a frame to place around the picture.
Persistence may make non-participation more socially costly, as it is more apparent when a user has not taken part in the cascade.
\item In a {\em nomination protocol}, a participating individual nominates one or more others to participate in the cascade, and the nominees, if they accept the challenge, will nominate others.
The nomination protocol usually demands a unique contribution requiring varying degrees of effort, and also introduces a social cost to not having one's nomination accepted or not accepting the nomination.
The Ice Bucket Challenge follows this protocol.
\item In a {\em volunteer protocol}, a participant asks for volunteers of others interested in joining, and subsequently assigns a task to be fulfilled by each of them, including calling for additional volunteers.
The effort required can vary, with lower social costs than nomination cascades, as only those who are interested volunteer to participate.
The music challenge meme described above follows this protocol.
\end{itemize}
Although we expect that these classes of protocols will not be exhaustive, they demonstrate both a substantial range of properties and a remarkably consistent relationship between the number of individuals who could participate at each step (\emph{exposed individuals}) and the fraction of those individuals who do participate in the spreading behavior (\emph{adopters}).

\xhdr{Summary of findings}
Our findings show how these four protocols lead to differences in the resulting cascade structure and growth.
As the individual effort required to participate increases, from transient and persistent copying to nomination and volunteering, the speed at which information propagates from one node to another decreases.
Though cascades with copy protocols (or \emph{copy cascades}) tend to rely on network hubs such as pages to spread, cascades with nomination or volunteer protocols (or \emph{nomination cascades} and \emph{volunteer cascades} respectively) tend to do so primarily via lower-degree nodes, from person to person.
As the social cost of not participating increases, the more likely it is that a cascade will spread through strong ties (e.g., when two friends have many friends in common).
Higher social cost also tends to result in cascades exhibiting complex diffusion, suggesting that participants wait to observe the behavior of others before acting.
We also demonstrate how the predictability of transmission varies across these different diffusion protocols; predicting successful transmission is easier for nomination cascades than for copy cascades.

Regardless of diffusion protocol, we find that the most successful cascades exhibit a reproduction number close to 1.8 -- that is, each new participant results in 1.8 additional participants on average throughout its life -- suggesting that successful protocols balance audience size and likelihood of participation.
The two copy protocols have a large potential audience at each step but low likelihoods of participation, while the nomination and volunteer protocols have a smaller number of potential participants at each step but a higher likelihood of any individual participating.
It is striking that across all these types of cascades, the most successful achieve a synthesis of these factors leading to a consistent maximum reproduction number.

Last, by studying the structure of the resulting cascades, we show how observing even a small fraction of a cascade's structure allows us to differentiate between diffusion protocols and how cascades with these protocols may be generatively modeled.

In summary, we:
\begin{itemize}[noitemsep]
\item identify four diffusion protocols for cascade propagation;
\item show how these diffusion protocols can be characterized in terms of individual effort and social cost and study how transmission predictability varies by protocol; and
\item demonstrate how the structure of the resulting cascades can differentiate these protocols and also be modeled.
\end{itemize}

\section{Background}

\xhdr{Cascades}
Most relevant to the present work is research that has investigated the network structure of the information diffusion cascades enabled by sharing features in online social networks~\cite{cha2009measurement,lerman2010information}.
These studies have shown how such cascades vary in depth, from the relatively shallow and broad cascades of Twitter~\cite{goel2012structure} and online games~\cite{bakshy2009social} to the deep chains formed by email petitions~\cite{liben2008tracing} and copy-and-paste memes~\cite{adamic2016information}.
Prior literature has also quantified social influence in online information sharing
\cite{cha2010measuring,bakshy2011,romero2011influence,aral2012identifying} and studied the predictability of a cascade's size and depth \cite{cheng2014can,khosla2014makes}.
This body of work has typically focused on simple information-replicating cascades (e.g., resharing or retweeting), generally arguing that hubs are crucial to the success of these cascades \cite{wasserman1994social,cheng2014can}.

Still, other cascading mechanisms for diffusion exist beyond those where information is simply broadcast.
For instance, cascades may spread in a targeted fashion.
Word-of-mouth marketing uses personal referrals in lieu of advertising \cite{buttle1998word}, and research has examined viral marketing mechanisms where purchasers recommend products to specific friends to receive discounts \cite{leskovec2007dynamics}.
Transmitted information may also persist and confer social identity, e.g., in the case of ``equals sign'' profile pictures on Facebook~\cite{vie2014defense}.
And rather than a format of passively receiving and passing on information, cascades may involve back-and-forth communication, as in the case of the self-organization of online volunteer efforts during natural disasters \cite{starbird2011voluntweeters}.
This diversity of mechanisms that prior literature has collectively considered motivates the present work on diffusion protocols.
We contribute a comparative focus to this body of work by examining the interplay between these protocols and the growth of the resulting cascades.

\xhdr{Individual effort and social cost}
Prior work studying social influence and collective action suggests that the individual effort and social costs associated with the mechanism of participation may be useful in predicting subsequent participation, and thus the growth of a cascade.

Models of collective action usually account for the resources or effort expended by individuals in deciding to contribute to the collective good, where individuals only participate if they gain more than they expend \cite{marwell1988social}.
In fact, lower individual costs (e.g., in required skill) can increase participation in collective action \cite{oliver1984if}.
These findings suggest that greater individual effort required to take action will reduce the probability of adoption.

At the same time, the likelihood of participating increases with the number of other people who have already adopted (i.e., with increasing social proof), as people seek to align their actions with what others have deemed as appropriate \cite{granovetter1978threshold,cialdini1993psychology}.
Costs to non-adoption can also exist.
When individuals are explicitly called upon to act in a certain way and fail to do so, they may lose face in front of meaningful others \cite{goffman1967face,ho1976concept}.
In other words, increased social cost associated with inaction, or conversely, greater social benefit associated with action, can increase the likelihood of adoption.

Further, the strength of the relationship between the required social proof and adoption likelihood increases with the perceived costs, whether direct (e.g., the cost of a fashion item), or indirect (e.g., in suddenly ceasing to participate) \cite{granovetter1978threshold}.
Overall, the diversity of mechanisms of social diffusion created in the online world calls for a more precise examination of the relationship between the mechanisms and the expected features of the generated social cascade.

\section{Diffusion Protocols}

While many existing models presume that information is transmitted between actors in a single atomic action, the particular protocol of transmission may involve multiple steps.
By understanding the diverse diffusion protocols that generate cascades, we may better understand (and potentially predict) the growth and resulting structure of a cascade.

\begin{figure}[t]
  \centering
  \includegraphics[width=1\columnwidth]{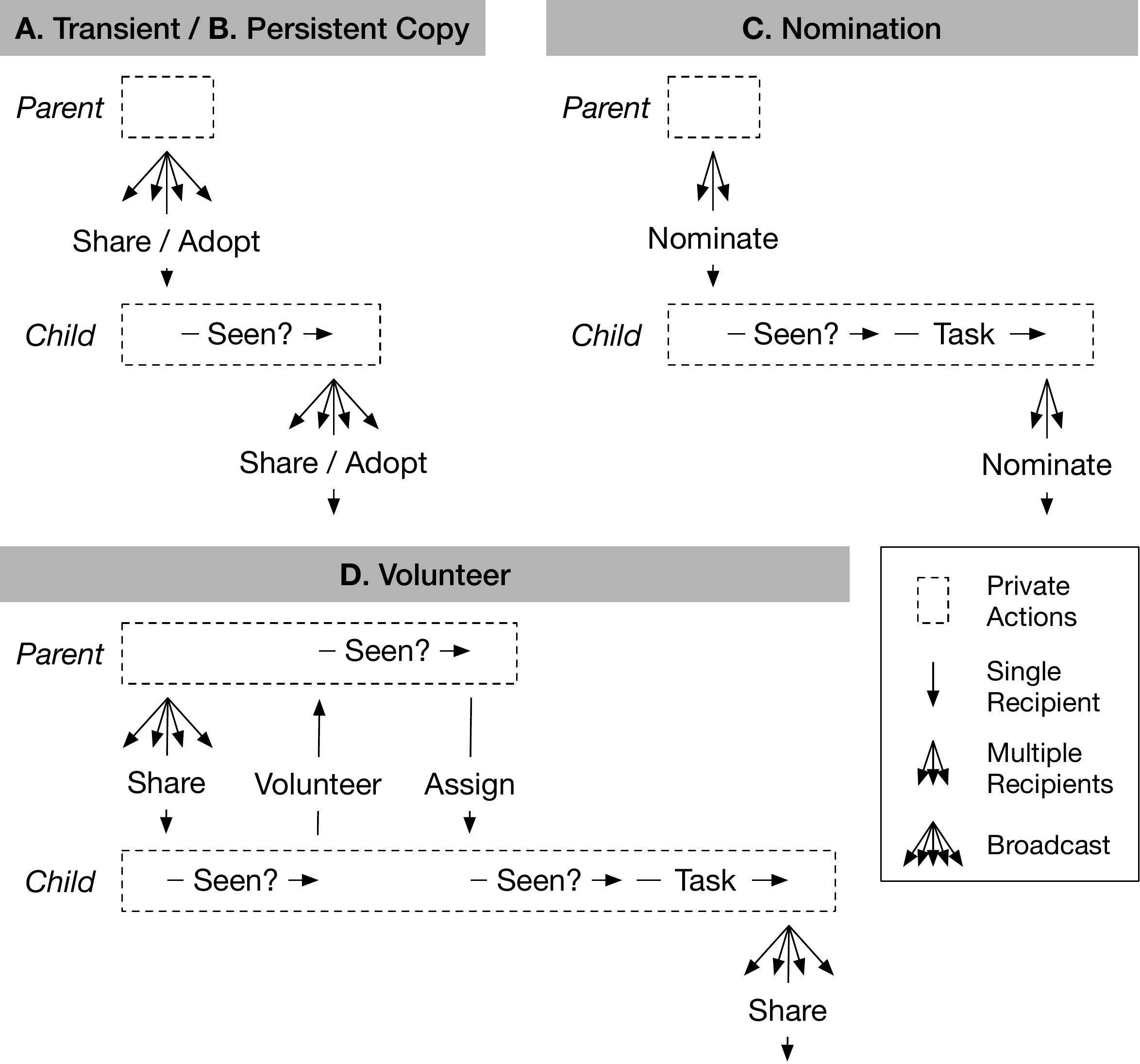}
  \caption{We identified four diffusion protocols: (a) transient and (b) persistent copy protocols, (c) nomination protocols, and (d) volunteer protocols. Each protocol describes a sequence of actions that results in successful propagation. While copy protocols involve users broadcasting content to all their friends, volunteer and nomination protocols require back-and-forth communication between users.}
  \label{fig:cascade_types}
\end{figure}

\subsection{A Vocabulary for Diffusion Protocols}

In this work, we identify and examine four protocols of transmission that exist on social media.
These four protocols underlie the largest observed cascades on Facebook (Fig.~\ref{fig:examples}) and, in representing different combinations of individual effort and social cost, allow us to understand how these factors influence cascade growth.

\emph{Transient} and \emph{persistent copy protocols} describe the diffusion mechanisms when content is simply replicated in a network.
Consider how a link or image may be shared from person $A$ to person $B$.
Fig.~\ref{fig:cascade_types}a illustrates this transmission as a three-step process: $A$ first shares the link with all of her friends, $B$ sees the link, and then $B$ decides to share the link with his friends in turn.

If the content in a copy cascade is only visible to others for a short period of time (e.g., when a person views a shared news article in their Facebook news feed), we call the protocol \emph{transient}.
In contrast, the content in a \emph{persistent} copy protocol is visible for an extended period of time (e.g., when individuals change their profile pictures to include a pride flag;~\citealp{adamic2015diffusion}).
As we later show, this distinction of persistence is important as it leads to cascades with substantially different growth patterns.

\emph{Nomination protocols} describe the diffusion mechanisms where transmission can only occur on edges explicitly selected by the sender and called out to the receiver.
The ALS Ice Bucket Challenge cascade is an example.
Instead of $A$ broadcasting the challenge to all of her friends, $A$ names a set of friends to challenge, which may include $B$.
For $B$ to propagate this cascade, $B$ must be included in that set of friends, see the challenge, complete the challenge (i.e., pouring a bucket of ice water over his head), and then name a new set of friends to challenge in turn (Fig.~\ref{fig:examples}c, \ref{fig:cascade_types}c).

\emph{Volunteer protocols} describe a mechanism where successful propagation requires two-way communication between participants.
For the music challenge meme cascade, $A$ first asks all her friends to ``like'' her post if they are interested in participating. $A$ then assigns a letter of the alphabet to each liker in a comment. Each assigned liker must then post about a musician whose name starts with that letter and replicate the instructions for further transmission (Fig.~\ref{fig:examples}d, \ref{fig:cascade_types}d).
That is, volunteer cascades consist of an invitation to action, interested individuals signing up, the assigning of tasks to each individual, the tasks' completion, and, finally, each individual broadcasting an invitation to action to others in turn.

One can imagine formulating new protocols by mixing and matching various components of these four types.

\begin{table*}[t]
  \centering
  \ra{1.3}
  \begin{tabularx}{\textwidth}{X|l|l|l|l}
  \toprule
    & \textsc{\footnotesize Transient Copy} & \textsc{\footnotesize Persistent Copy} & \textsc{\footnotesize Nomination} & \textsc{\footnotesize Volunteer}  \\ \hline
    Number of Cascades & 34 & 56 & 5 & 3 \\
    Mean Adoptions & 1.01 $\times$ 10\textsuperscript{6} & 8.88 $\times$ 10\textsuperscript{5} & 6.06 $\times$ 10\textsuperscript{6} & 9.91 $\times$ 10\textsuperscript{5} \\
    Mean Exposures Per Adopted Individual & 79.2 & 160 & 5.84 & 51.7 \\
    \% Shares from Top 1\% of Individuals & 13.7\% & 37.3\% & 3.29\% & 7.57\% \\
    Mean Mutual Friends & 28.8 & 70.7 & 58.1 & 44.6 \\
    Mean Prior Exposures & 1.25 & 3.87 & 1.33 & 1.22 \\
    Mean Adoption Delay (s) & 22.5 & 153 & 4.42 $\times$ 10\textsuperscript{4} & 1.31 $\times$ 10\textsuperscript{4} \\
    \bottomrule
  \end{tabularx}
  \caption{Properties of the cascades that result from each diffusion protocol. While each protocol can generate large cascades, the resulting cascades differ structurally and temporally.}
  \label{tab:data}
\end{table*}

\xhdr{Discussion}
Relative to copy cascades, nomination cascades and volunteer cascades have an explicit call to action and involve a direct request from one user to another.
This explicit request introduces a cost of non-compliance, that of \emph{losing face}, or the loss of social value by not taking a particular action \cite{goffman1967face}.
Once tagged in a nomination cascade, the choice for an individual is to follow through with the request or risk others noticing that they have not done so.
Importantly, such a face-losing cost can only be imposed if there is perceived widespread agreement on desirability of the behavior.
The ALS cascade is one such example: few would disagree that helping to fight Lou Gehrig's disease is generally desirable.
In contrast, when sharing a funny meme, there is still an expectation that friends may reshare it, but it is hard to imagine that someone would be held accountable for not doing so.

This face-losing cost associated with non-compliance thus allows for a larger cost associated with compliance (i.e., requiring greater individual effort).
All else equal, a copy cascade associated with ALS would likely have been less successful at generating donations without a clear social enforcement mechanism.
However, the very face-related mechanism that ensures the effectiveness of challenge cascades is also responsible for their rarity.
When an individual fails to complete a challenge, this not only translates into lost face for that individual, but also results in a status loss for the challenger.
This makes challenges risky for both the one challenged and the one challenging.
A fruitless challenge can be interpreted not only as a lack of interest on part of the challenged individual, but also as a lack of influence on part of the challenger.

The risk of losing face, in the context of an explicit call to action, can nonetheless be mitigated by mechanisms that allow one to instead save face, or discharge one's perceived obligation.
Unsurprisingly, the most common form of the ALS challenge included the option of donating as an alternative to the ice-cold shower.

Social proof can also make it more likely for individuals to take similar action when others have already done so \cite{sherif1935study}.
As participation in persistent copy cascades is more visible to others longer than in transient copy cascades, individuals may be more likely to participate to conform to the behavior of others, especially if they are members of the same social group.

Additionally, a ``macro-level'' diffusion pattern~\cite{strang1998diffusion} is likely to exist \textit{across} cascades, rather than within them.
There are only a few major types of information cascades on Facebook, with most using the same relatively narrow vocabulary of diffusion protocols -- copy protocols being common and volunteer and nomination protocols constituting distinctive niches.
This limited range of common replication strategies can be explained using the theory of \textit{mimetic isomorphism}, or how actors copy established practices in situations of diffusion under uncertainty \cite{meyer1977institutionalized}.
This theory has been used to explain the diffusion of successful tactics from one social movement to another \cite{strang1998diffusion}.
That individuals participating in new informational cascades on Facebook are likely to copy already-proven tactics (e.g., a ``reshare if you agree'' message) appears thus to be a logical extension of this theory to online platforms.

\subsection{Data}
We analyzed 98 cascades that occurred on Facebook from mid-2014 to early 2016.
These cascades generated over 117M posts which were viewed over 4B times by over 200M individuals.
All data was de-identified and analyzed in aggregate.
Each cascade is a directed acyclic graph ${G=(V,E)}$, where $V$ is the set of participants, and $E$ is the set of parent-child relationships in the diffusion process.
Only an individual's first participation in a cascade was considered, with subsequent actions merged with the first.
If a user interacts with more than one previous post in the cascade, we preferentially select the first one that user saw as the ``parent'' of that user's post.

We sampled cascades spread via each protocol, using de-identified data processed by automated scripts, as follows:
\begin{itemize}
\item {\bf Transient and persistent copy cascades}. We sampled 34 of the largest reshare cascades by number of shares from January 2016, as well as 56 of the largest profile-picture-frame-adoption cascades from the same time period, tracking them for the subsequent 28 days.
\item {\bf Nomination cascades}. We searched for posts containing words relating to nomination (e.g., ``nominate'' and ``challenge''), and that mentioned one or more individuals, and identified five different nomination cascades, including the ALS Ice Bucket Challenge.
We used an SVM classifier to distinguish nominators and nominees (e.g., Bob and Carlos in Fig.~\ref{fig:examples}c; AUC $\ge$ 0.88 for each cascade).

\item {\bf Volunteer cascades}. We searched for posts containing words such as ``assign'' (since participants are assigned to complete specific tasks) and ``fill'' (e.g., the music challenge meme cascade asked participants to ``fill Facebook with music'', but past cascades have asked participants to fill or occupy Facebook with everything from art to poetry to Nicolas Cage), and identified three different large volunteer cascades.
\end{itemize}

Our sample has fewer nomination and volunteer cascades because they occur less frequently than both types of copy cascades.
Generalizing our findings on the nomination and volunteer cascades in this sample to a broader set of such cascades remains future work.

\section{Properties of Different Diffusion Protocols}

\begin{figure*}[tb]
  \centering
  \begin{subfigure}[b]{0.33\textwidth}
    \includegraphics[width=\textwidth]{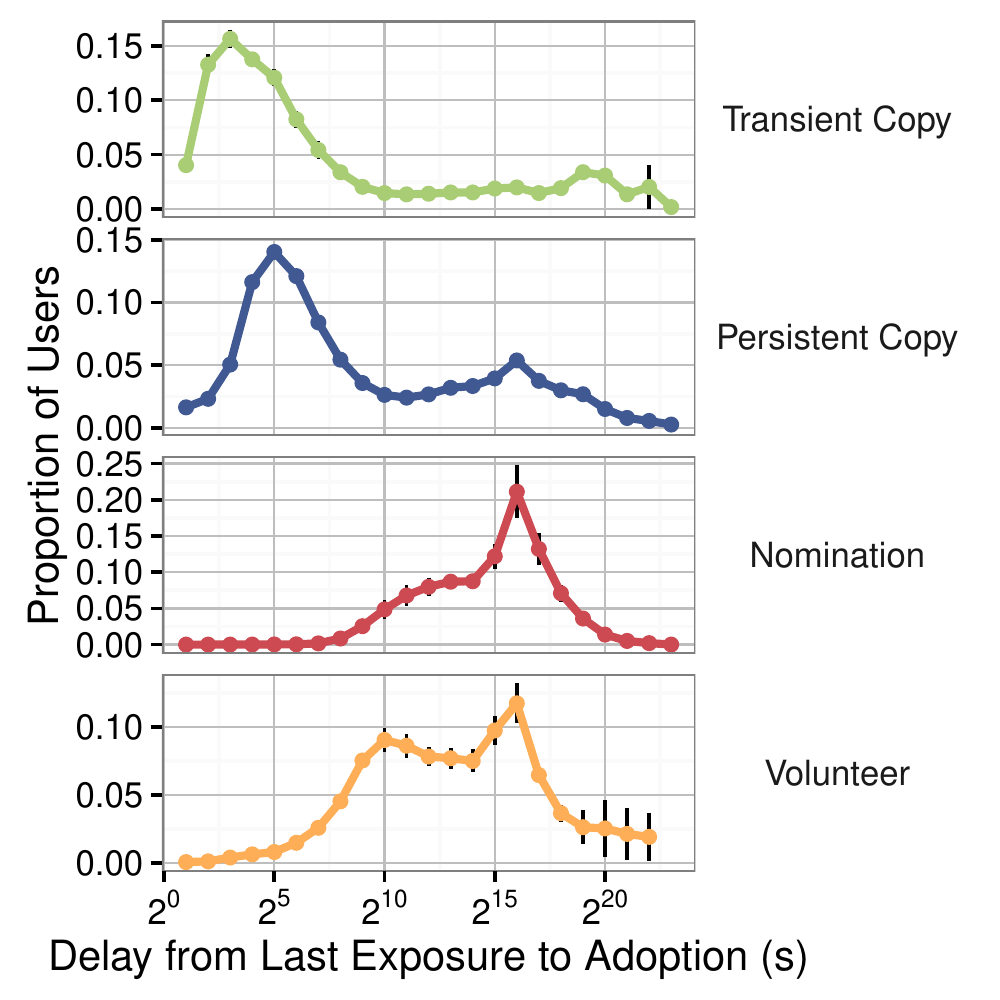}
    \caption{}
    \label{fig:exposure_stats_1}
  \end{subfigure}
  \begin{subfigure}[b]{0.33\textwidth}
    \includegraphics[width=\textwidth]{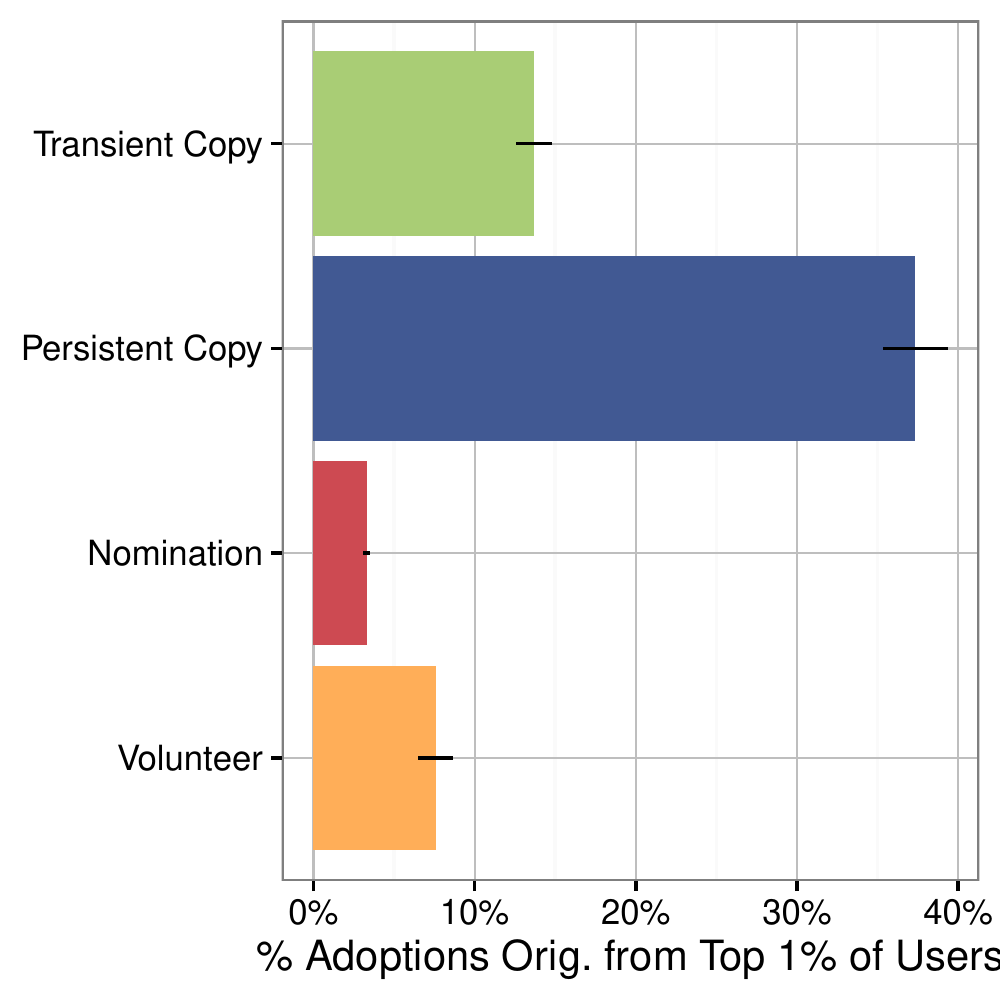}
    \caption{}
    \label{fig:exposure_stats_5}
  \end{subfigure}
  \begin{subfigure}[b]{0.33\textwidth}
    \includegraphics[width=\textwidth]{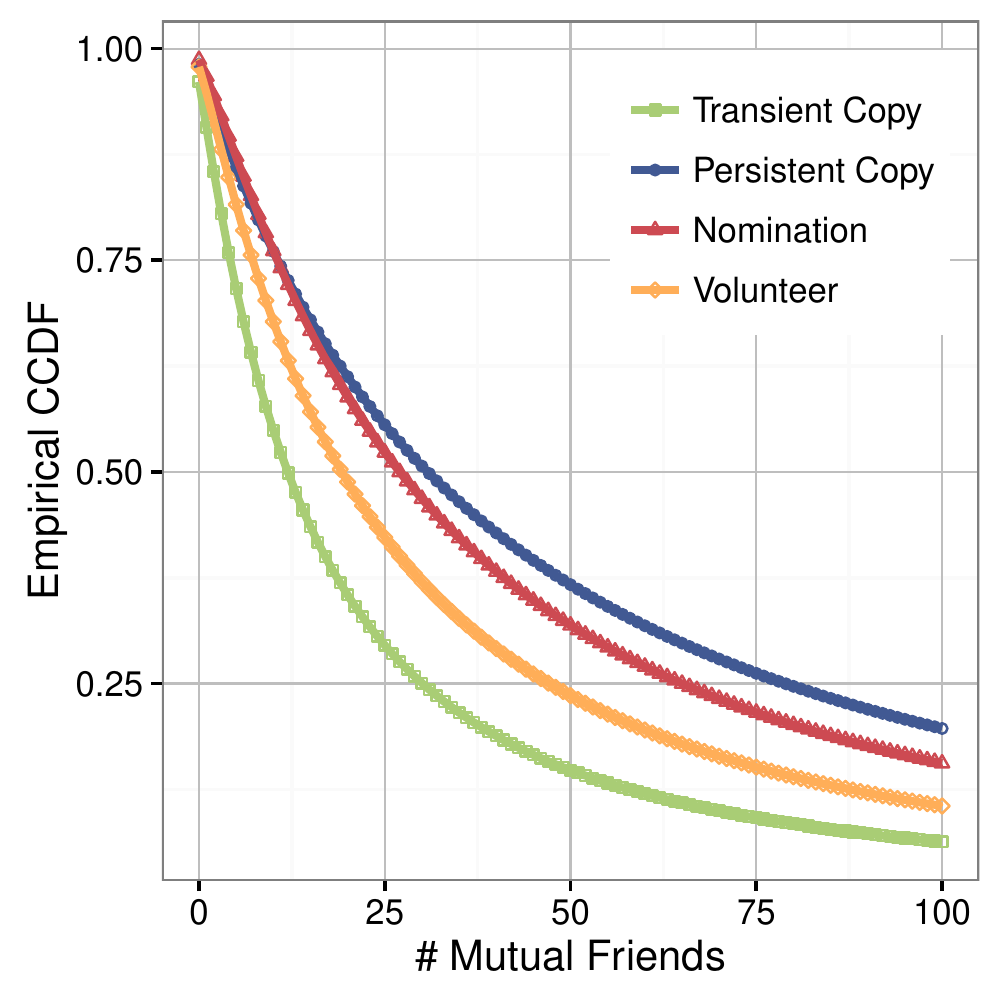}
    \caption{}
    \label{fig:distributions_1}
  \end{subfigure}
  \caption{(a) Copy cascades are transmitted from individual to individual more quickly than volunteer or nomination cascades. (b) Copy cascades are also more reliant on hubs (i.e., nodes with high degree) for distribution, while volunteer and nomination cascades spread more evenly. (c) However, while transient copy cascades spread through weaker ties, persistent copy cascades spread through stronger ties. (Error bars denote standard errors of the mean across cascades.)}
  \label{fig:observations_1}
\end{figure*}

\begin{figure*}[tb]
  \centering
  \begin{subfigure}[b]{0.33\textwidth}
    \includegraphics[width=\textwidth]{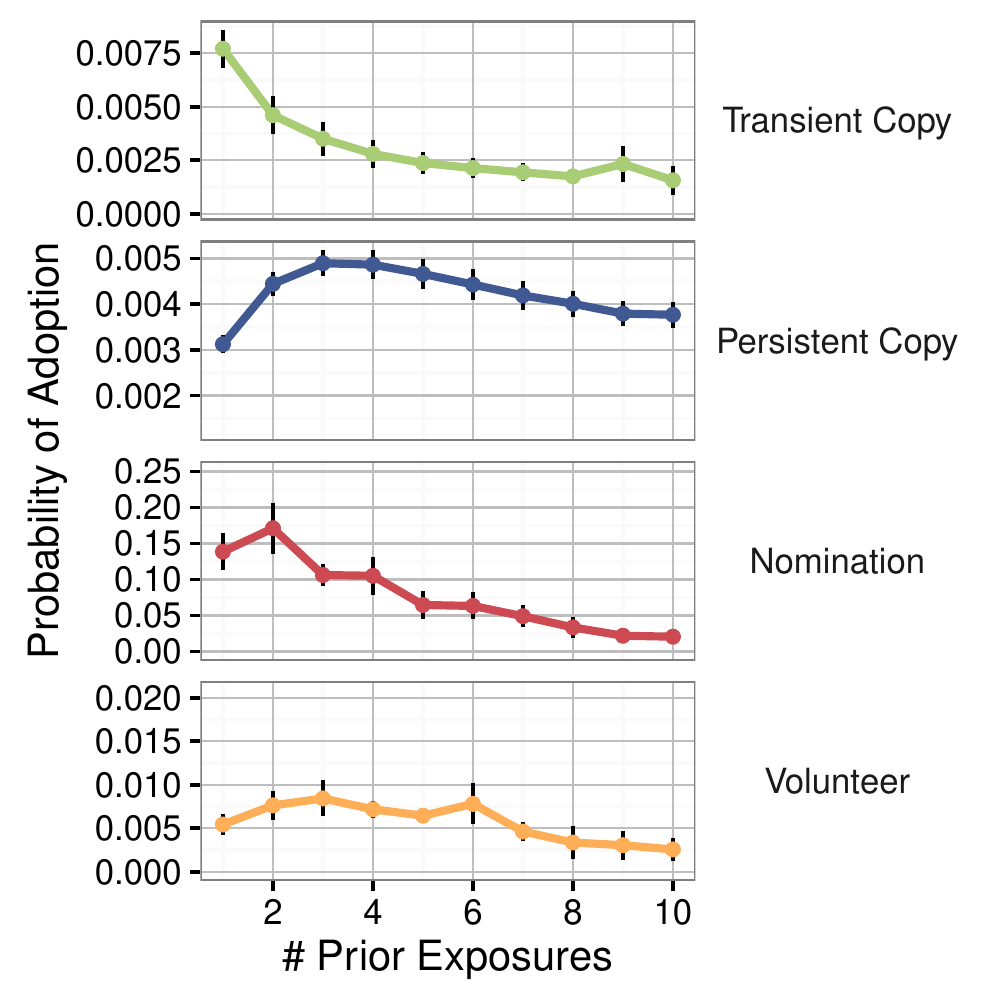}
    \caption{}
    \label{fig:exposures_1}
  \end{subfigure}
  \begin{subfigure}[b]{0.33\textwidth}
    \includegraphics[width=\textwidth]{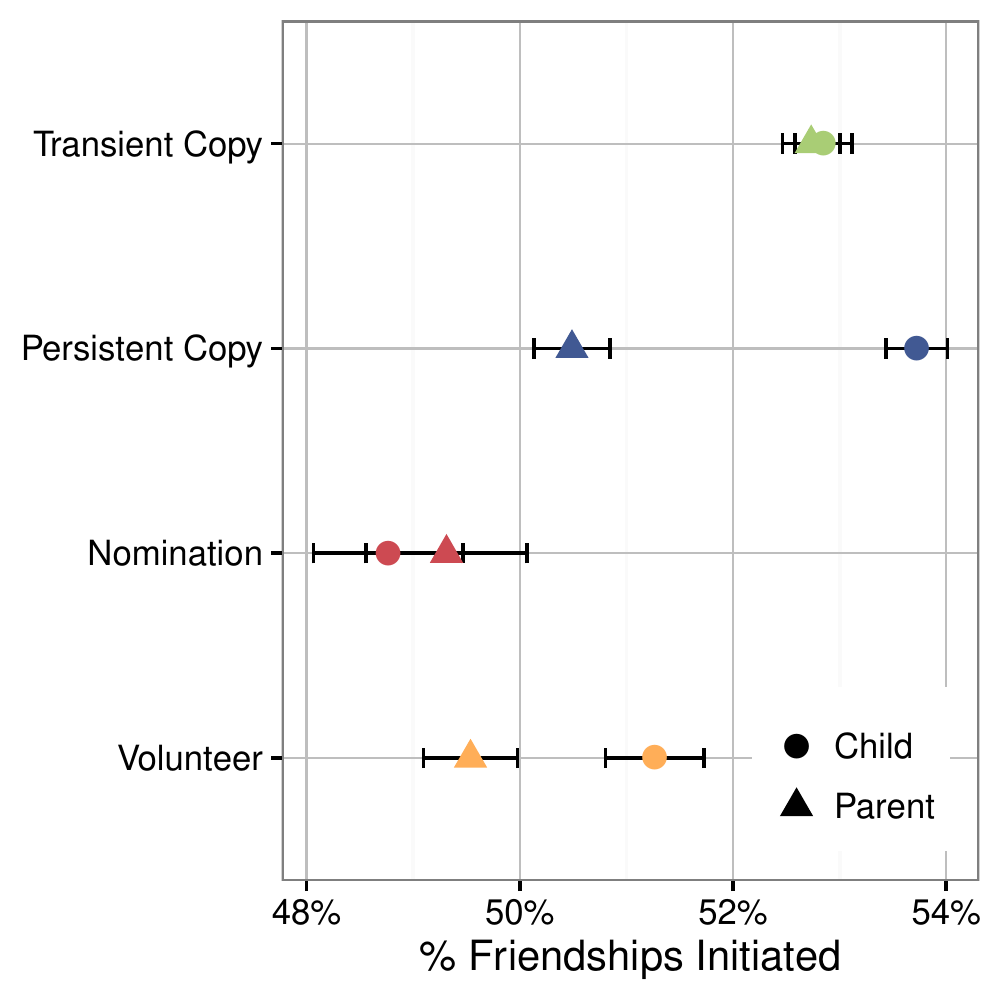}
    \caption{}
    \label{fig:status_4}
  \end{subfigure}
  \begin{subfigure}[b]{0.33\textwidth}
    \includegraphics[width=\textwidth]{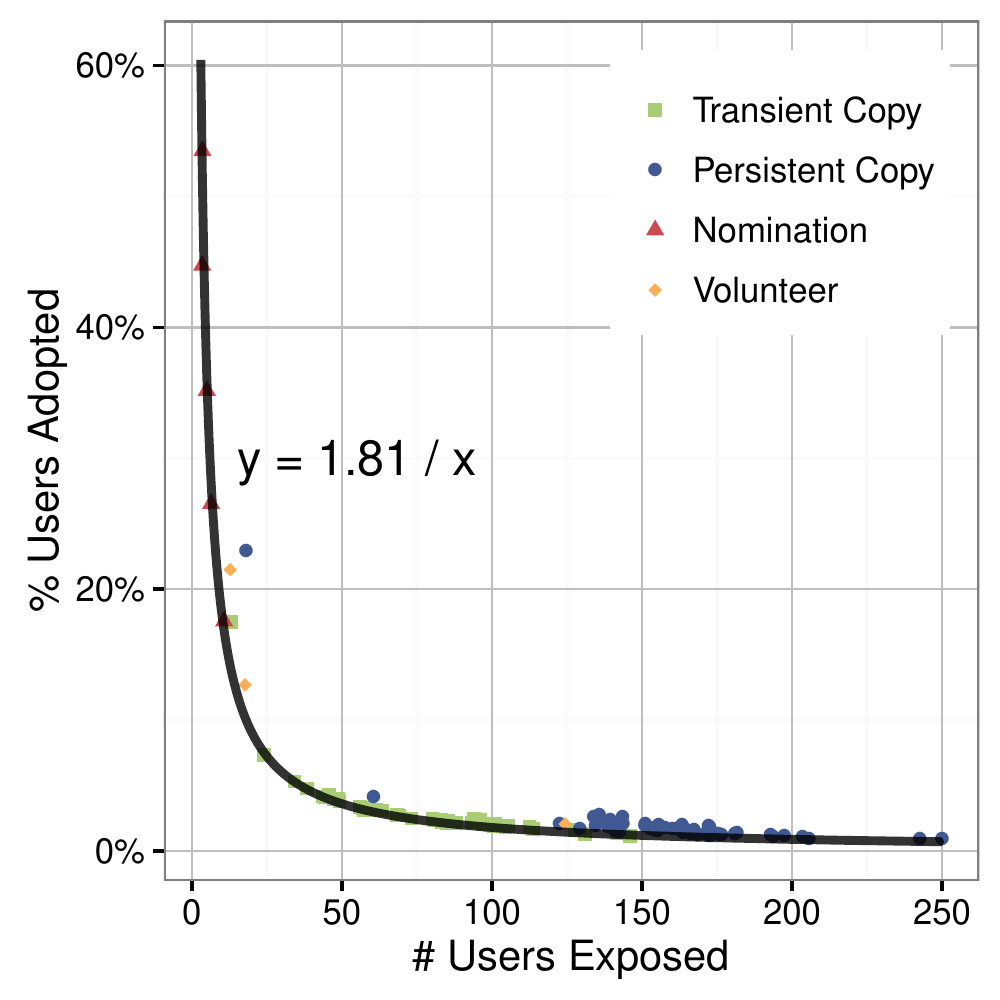}
    \caption{}
    \label{fig:exposure_stats_3}
  \end{subfigure}
  \caption{(a) Transient copy and nomination cascades exhibit properties of simple diffusion, while other types of cascades exhibit properties of complex diffusion. (b) Persistent copy and volunteer cascades tend to flow from higher-status to lower-status individuals, while transient copy and nomination cascades do not. (c) Regardless of the transmission protocol, each additional adoption results in an average of 1.81 subsequent adoptions.}
  \label{fig:observations_2}
\end{figure*}

Here, we detail how the cascades that result from each diffusion protocol differ on several important axes.
We also show how these observations arise from differences in the individual effort required to participate in a cascade and the social costs associated with non-participation.

\subsection{Propagation Speed}
As the complexity of a diffusion protocol and effort demanded on each participant's part increases, we expect an increase in the time taken for participation to be transmitted across an edge.
Transient and persistent copy cascades, whose protocols require the least participant effort, spread more quickly (with a mean delay of 23 seconds and 2.6 minutes respectively), while volunteer and nomination cascades, which involve more steps and the completion of a task, spread more slowly (3.6 and 12.3 hours, respectively; $p$$<$0.01 using Mann-Whitney tests with a Holm correction; Fig.~\ref{fig:exposure_stats_1}).

\subsection{Network Hubs}
In most work on cascades, hubs (or highly connected individuals) play a significant role in spreading information to their many social ties \cite{wasserman1994social,cheng2014can}, with degree centrality commonly used as a measure of influence \cite{goldenberg2009role}.
As shown in Fig.~\ref{fig:exposure_stats_5}, the top 1\% most-connected users or pages were responsible for a much larger fraction of subsequent adoptions in copy cascades than in other cascades ($W$$=$750, $p$$<$0.01).
These findings corroborate other work suggesting that the strength of hub nodes lies in informing but not persuading \cite{delre2010will}, and that protocols requiring greater individual effort may not benefit as much from hubs; the more friends one has, the less influence one has over each friend \cite{katona2011network}.
In other words, rather than spreading successfully through wide distribution, other cascade types tend to spread more from person to person, suggesting that as individual effort and social costs increase, hubs become less effective at motivating participation.

\subsection{Tie Strength}
Differences in tie strength between individuals -- \emph{strong} ties between close friends, vs.~\emph{weak} ties between acquaintances -- may explain hubs' ineffectiveness in certain cases.
One proxy of tie strength is the number of mutual friends two individuals share \cite{shinetworks}.
We find that transient copy cascades tend to spread through weaker ties with fewer mutual friends on average than other cascades types (28.8 vs. $\mathop{\ge}$44.6, $W$$>$7, $p$$<$0.05), which require either more individual effort or impose greater social cost.
Given tie strength's correlation with geographic proximity, we also find that ties in persistent copy cascades are somewhat closer on average (653 km vs.~$\mathop{\ge}$756 km apart, n.s.).

While the degree of nodes can vary considerably, the number of strong ties tends to be bounded by the effort required to maintain the tie~\cite{dunbar2012social}.
Where little effort is required and information flows freely, such as in copy cascades, hubs play an outsized role in furthering the spread of the cascade.
Their role is reduced, however, if the cascade spreads primarily through strong ties, where the number of such ties per node is much more evenly distributed.

But why then do persistent copy cascades, which are just as simple as transient copy cascades in requiring low individual effort, spread through much stronger ties than any other type of cascade?
The persistence of these cascades suggests that they also contribute to one's identity in relation to a group.
In such a case, one would be more likely to participate in a cascade if many friends are also doing so to demonstrate group belonging.
This idea is consistent with many of the profile picture frame cascades we observed identifying the individual as a fan of a sports team.

\subsection{Complex Diffusion}
As tie strength is of variable importance in the transmission of different cascades, cascades of different protocols may spread either through communities or across communities: strong-tie cascades may spread within communities, while weak-tie cascades may easily cross from one to another.
To this end, we examined the median number of friends that participated in a cascade prior to each participant joining as a proxy for whether a cascade tends to ``jump'' to a different part of the network with each transmission.
Corroborating our previous findings on tie strength, we find that this value was highest for persistent copy cascades (8.29), followed by nomination (6.30), volunteer (3.07), and transient copy cascades (1.54).

The observation that an individual adopts after several others can not only be reflective of an underlying community structure, but also of complex diffusion that requires multiple exposures for adoption \cite{centola2007complex,adamic2015diffusion}.
Several confounds exist, and to differentiate between these two scenarios, we need to account for whether an individual actually saw a friend's participation.

Fig.~\ref{fig:exposures_1} illustrates how the likelihood of adoption (or participation) changes with the number of exposures (or views of the information transmitted in that cascade).
Comparing the adoption likelihood after one exposure with that after two exposures, we find that the adoption likelihood is lower after two exposures for all transient copy cascades and all but one nomination cascade, but higher for all but two persistent copy cascades and all volunteer cascades ($\chi^2$$>$42, $p$$<$0.01).
This suggests that in the latter cascades, people tend to wait to observe the behavior of others before deciding whether to act~\cite{adamic2015diffusion}.
For profile-picture-frame adoption, this may mean waiting to see whether a community (e.g., fans of a sports team) is collectively adopting.
For volunteer cascades, due to the additional effort required to participate, individuals may wait to see if others are expending the effort.

Notably, nomination cascades appear to spread like simple diffusion -- while each nominee was explicitly called out, the greatest social value is likely in responding earliest (i.e., in appearing to be most enthusiastic to both nominator and friends).
A volunteer's dilemma \cite{diekmann1985volunteer} may also contribute to this effect:
a nominator will look bad if no one they nominate participates, so at least one nominee may end up participating to avoid this negative outcome.

\subsection{Parallels to Friending Behavior}
The friendship network underlying cascades is also formed through an invitation mechanism, where one friend invites another to form a friendship tie.
Here, we compare individuals' tendency to initiate and form social ties with their role in various cascade types.
We measure the percentage of a person's friendships on Facebook initiated by that person, as well as their total number of friends.
These quantities may be correlated with status.
Since connections with high-status individuals lead oneself to be viewed more positively \cite{sauder2012status}, a lower percentage of friendships initiated may correlate with higher status, and, similarly, a larger number of connections with higher status.
Still, these quantities may also correlate with overall Facebook activity, and teasing these factors apart remains future work.

There is a mild differential for persistent copy cascades, where child nodes in the cascade initiate 53.7\% of friendships on average, as opposed to parent nodes (50.5\%) or those who saw the cascade but decided not to participate (50.6\%, $W$$\ge$0, \textit{p}$<$0.01; Fig.~\ref{fig:status_4}).
Volunteer cascades exhibit a similar but weaker effect.
In cascades with nomination protocols, the person receiving the nomination initiated a smaller proportion of their friendship ties (47.3\%).
If initiating fewer friendships does indeed correlate with higher status, then this implies a slight effect in which copy and volunteer cascades, where anyone is free to participate, tend to diffuse from higher to lower status participants. On the other hand, nomination cascades, where particular friends are selected to participate, tend to consistently diffuse via slightly higher-status individuals.

We see empirically similar results regarding an individual's number of friends.
In persistent copy cascades, existing participants have an average of 291 more friends than new participants, but this gap is smaller for other cascade types ($\mathop{\le}$178, $W$$\ge$87, $p$$<$0.05), reiterating the varying importance of high-degree hubs.

\subsection{The Constant Reproduction Number}
Regardless of diffusion protocol and the number of individuals exposed, and considering only the internal nodes of a cascade, we found that all cascades we observed exhibited a reproduction number close to 1.8 (Figure \ref{fig:exposure_stats_3}).
While this number varies over the life of a cascade, this overall consistency is remarkable, suggesting that lower exposure rates are offset by higher per-exposure adoption rates in successful cascades.
Copy cascades
have very large potential audiences in a user's network neighborhood, although there is little incentive for any individual to participate.
In contrast, nomination and volunteer cascades require more effort to participate and have smaller potential audiences, but their protocols place social pressure on each potential individual to participate (e.g., via the cost of losing face).

These observations parallel the different selective pressures suggested by $r/K$ selection theory \cite{pianka1970r}, used to understand evolutionary strategies. $r$-selected species emphasize high growth rates at the expense of individual survivability, while $K$-selected species invest more heavily in fewer offspring.

The constant reproduction number also points to a bystander effect: each exposed individual benefits from the participation of just a few, which then absolves others of the need to participate \cite{rutkowski1983group}.
Reproduction may also be limited by network saturation: people are less likely to share a link knowing that many of their followers have already seen it \cite{gilbert2012designing}.

\subsection{Predicting Adoption}
How might the diffusion protocol that a cascade uses for its propagation affect its predictability?
Here, we consider the task of predicting whether an exposed individual will become an adopter.
For each cascade, we sampled pairs of individuals where both individuals were exposed, but only one adopted the spreading behavior.
This results in a balanced prediction task where random guessing results in 50\% accuracy.

Table \ref{tab:performance_all}a summarizes the mean AUC obtained for each category using a random forest predictor on demographic and network features (e.g., age, gender, family ties, friend count, mutual friends, and geographic location).
We find that nomination cascades are most predictable, while persistent copy cascades are the least predictable.
Across all categories, mutual friend count was the strongest individual predictor of adoption (mean AUC$\mathrel{=}$0.66), followed by age (0.56) and friend count (0.56).
Notably, the strongest feature for persistent copy cascades was the number of friend initiations (0.57).

\subsection{Summary}
\begin{figure}[tb]
  \centering
  \includegraphics[width=1\columnwidth]{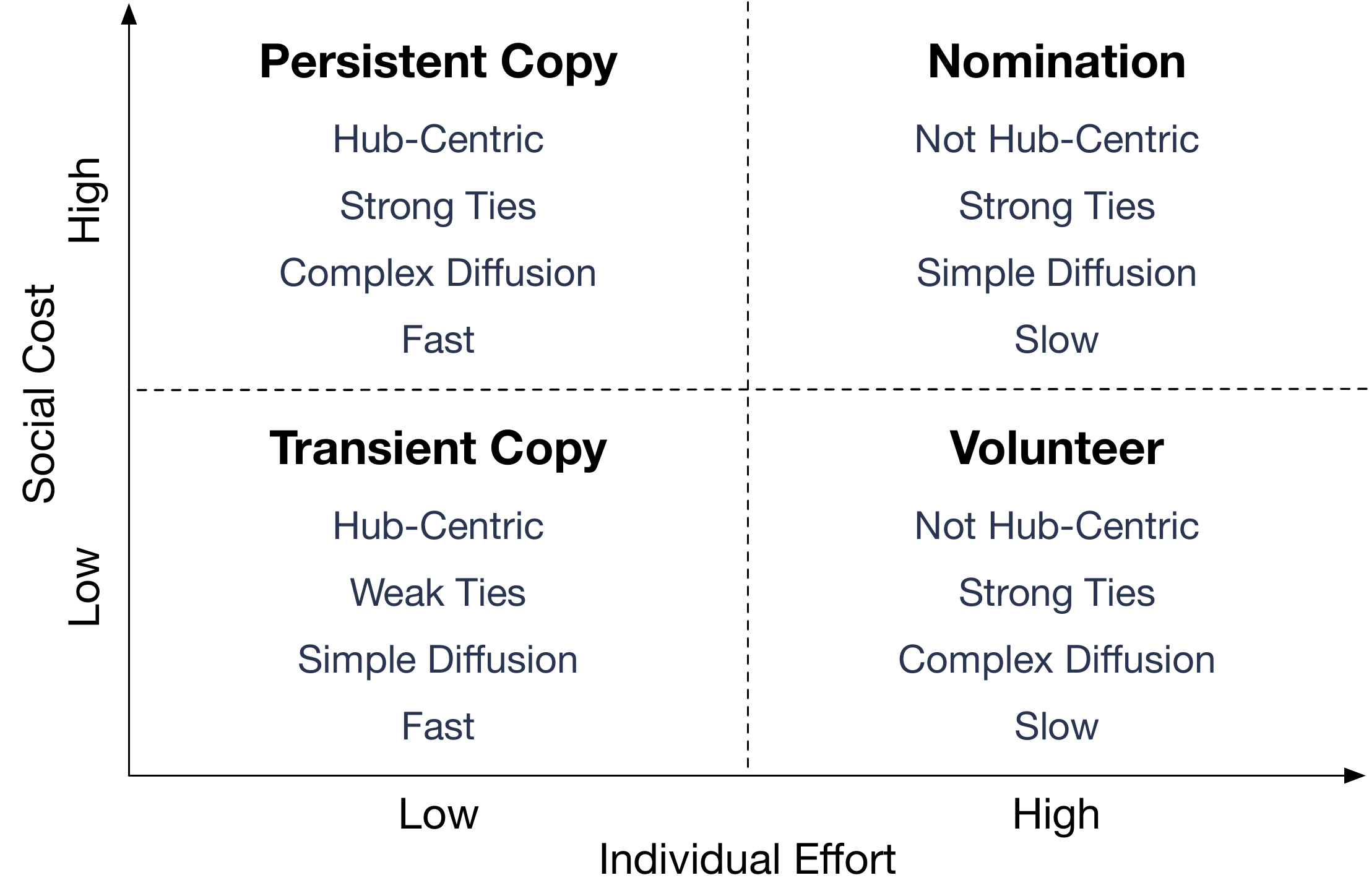}
  \caption{Overall, diffusion protocols differ primarily in terms of the individual effort required to participate, and the social cost of participation. Though each protocol can result in large cascades, the properties of the resulting cascades differ substantially.}
  \label{fig:axes}
\end{figure}

Figure \ref{fig:axes} illustrates how diffusion protocols may be organized along two axes: individual effort, or the work required to participate in a cascade; and social cost, or the social obligations and norms associated with participation.
When the individual effort and social cost involved in participating in a cascade are low, highly connected ``hub'' users are effective distributors.
When the individual effort required is high, the wide exposure these hub users enable becomes less important.
If the individual effort required or social cost incurred is high, strong ties become key to propagation as they create social pressure to participate.
When participation is voluntary and either significant effort or social cost is incurred, social influence leads to cascades exhibiting complex diffusion -- people observe others' behavior to decide whether to participate.
Finally, as the individual effort required to participate increases (e.g., requiring complex handshakes or completing difficult tasks), the speed of propagation decreases.

\section{Diffusion Protocols and Cascade Structure}

Thus far, we have shown how different diffusion protocols, by requiring varying levels of individual effort and social cost, can result in substantially different cascades.
We now turn to better understanding the resulting structures of these cascades through a series of prediction tasks.
In particular, we look at how a cascade's underlying diffusion protocol can be inferred through its structure, and how we may begin to use generative models to represent these cascades.

\begin{table*}[tb]
\centering
\ra{1.3}
\begin{tabular*}{\textwidth}{@{\extracolsep{\fill}}ll|p{1cm}p{1cm}p{1cm}p{1cm}|p{1.2cm}p{1.2cm}p{1.2cm}}\toprule
	\multirow{2}{*}{\textbf{Protocol}} & \multicolumn{1}{c|}{\multirow{2}{*}{\makecell{\textbf{(a)} Predicting\\Transmission}}} & \multicolumn{4}{c|}{\textbf{(b)} Differentiating Cascades} & \multicolumn{3}{c}{\textbf{(c)} Differentiating Real and Synthetic} \\
    & & {\small T. Copy} & {\small P. Copy} & \small{Nomin.} & \small{Volun.} & \small{Baseline} & \small{Degree} & \small{Cond.~Degr.} \\
	\hline
    Transient Copy & 0.64 & 0.55 & 0.64 & 0.61 & 0.68 & 0.83 & 0.53 & \textbf{0.51} \\
    Persistent Copy & 0.63 & - & 0.59 & 0.61 & 0.67 & 0.90 & 0.54 & \textbf{0.52} \\
    Nomination & 0.82 & - & - & 0.58 & 0.65 & 0.68 & 0.58 & \textbf{0.55} \\
    Volunteer & 0.75 & - & - & - & 0.59 & 0.84 & 0.58 & \textbf{0.51} \\
\bottomrule
\end{tabular*}
\caption{(a) It is easiest to predict whether a nominated individual is likely to participate in a nomination cascade (AUC $\mathrel{=}$ 0.82), and hardest to predict whether an exposed individual is likely to participate in a (transient or persistent) copy cascade (0.63).
(b) A classifier can differentiate subtrees of cascades spreading through different diffusion protocols, suggesting that a cascade's structure can predict its diffusion protocol.
(c) Comparing mean performance (AUC) across cascades on differentiating actual sampled subtrees from each cascade and synthetically-generated subtrees, the conditional degree model is best able to model these subtrees as a classifier performs worst at distinguishing these synthetic subtrees from the actual ones.}
\label{tab:performance_all}
\end{table*}

\subsection{Differentiating Protocols with Cascade Structure}
Beyond predicting transmission, we can also attempt to differentiate between the structures that result from different diffusion protocols.
In particular, is it possible to differentiate the adoption cascades of different diffusion protocols, even if we observe just a small fraction of the structure of each cascade?
In contrast to differentiating entire cascades from one another, where size alone can already be strongly discriminative, we expect that this is a more difficult task, given that we can only compare a small random fraction of an entire cascade to that of another cascade.

Specifically, we sample subtrees from two cascades (here, we consider a paired prediction task), with replacement, by selecting nodes at random, then traversing them until we have obtained subtrees of at most depth $d$.
Our task is then to predict which subtree came from which cascade.
This particular formulation is motivated by the fact that it may be difficult or even impractical to recover entire cascades, especially if they are continuing or recurring, but much easier to observe small parts of these cascades as they spread locally.
We sampled 400,000 subtrees from pairs of cascades (using $d$=3) and computed multiple structural features on these subtrees (e.g., number of nodes and edges, proportion of leaves, in-degree entropy) for prediction using a random forest classifier. (Note that we consider the edges of each tree as directed from children toward their parents; thus, a node's in-degree specifies the number of children it has.)

As shown in Table \ref{tab:performance_all}b, cascades of different diffusion protocols can be differentiated from each other (e.g., mean AUC$\mathrel{=}$0.68 differentiating a transient copy and a volunteer cascade), but difficult to tell apart cascades spreading through the same protocol (e.g., mean AUC$\mathrel{=}$0.55 telling apart different transient copy cascades).
Across all protocols, the strongest individual predictor was the average in-degree of the non-leaf nodes in the subtree (AUC$\mathrel{=}$0.59), or more simply, its branching factor.

\subsection{Modeling Cascades}
That the strongest individual structural feature that differentiates cascades of different diffusion protocols was their branching factor suggests that we may be able to model these subtrees using branching processes, which are commonly used to model epidemics \cite{allen2008mathematical}.
Here, we compare the effectiveness of several branching process models in accurately representing these cascades and consider three models of increasing complexity:
\begin{itemize}[noitemsep]
\item A baseline model, parameterized only by a branching factor $k$ and susceptibility $p$: each participant asks $k$ others to participate; each other participates with probability $p$.
\item A degree-based model, which generates child nodes using the expected degree distribution.
\item A conditional degree-based model, which conditions a node $u$'s indegree distribution on $u$'s parent's indegree.
\end{itemize}

To evaluate these models, we again sample 200,000 subtrees ($d$$=$3) for each cascade, and use these samples to fit the three branching process models.
We then generate 200,000 subtrees using each model, and then compare these synthetically generated subtrees against the original sampled subtrees.
We again model this as a binary prediction task, where a random forest classifier must differentiate a real subtree from a synthetic subtree using the structural features of the subtrees.
Here, the \emph{worse} the classifier performs, the better the model is able to approximate that cascade.

Performance is worst for the conditional degree model, suggesting that it best models the actual subtrees in these cascades.
Decreasing $d$ further reduces classifier performance (e.g., AUC$\mathrel{\approx}$0.50 for $d$$\le$2).
Even more complex models (e.g., conditioning on the $n$th sibling's degree distribution) resulted in empirically similar performance.
Altogether, these results suggest that regardless of diffusion protocol, branching processes can be used to model subtrees of these cascades.
Future work may involve considering other epidemic models (e.g., threshold models) or modeling other aspects of cascades (e.g., propagation speed).

\section{Discussion and Conclusion}

In this paper, we examined several classes of cascades, all of which were able to reach large size by spreading through Facebook. We demonstrated how understanding the underlying diffusion protocols, through the lens of participation effort and the social cost of non-participation, explains several important cascade features: size, speed, branching factor, and virality.

Cascades using protocols that require little effort and whose presence is transient tend to be adopted quickly, and typically after the very first exposure.
These cascades are most numerous and comprise a large fraction of the largest cascades.
It is common for a small proportion of the nodes to have influenced a significant fraction of adoption.
Because anyone observing the cascade can instantly participate, the greater the audience of a node, the more potential adoptions it can drive.
However, predicting which of the many exposures will lead to adoption is a difficult task.

In contrast, cascades using more effortful protocols tend to spread more slowly.
While such cascades may appear uncompetitive with the speed and broadcast mechanism of reshare cascades, a stronger interaction between influencer and influencee, via nomination or volunteering, compensates.
Because these cascades spread only through some ties rather than through broadcasts, their branching factor tends to be lower.
Interestingly, predicting whether someone who has been exposed in this type of cascade will adopt is easier than it is in the simple broadcast cascade.

We showed that these underlying protocols lead to sufficiently different structure that the protocol being used can be distinguished by cascade structure alone.
But despite the difference in the transmission mechanics and the resulting structure, the basic reproduction number of a range of cascades is 1.8, with higher adoption rates of more selective mechanisms compensating for the less exposure.

This work represents first steps in understanding the relationship between the diffusion protocols that describe a cascade's underlying spreading mechanisms and its overall growth.
While we have attempted to represent a diversity of protocols in our analysis, cascades other than the ones we studied, but that also follow these protocols may exist.
Other protocols may also arise over time, both on Facebook and on other platforms.
Adapting the effort and social cost framework we have constructed to capture this increased diversity remains future work.

While we sought to characterize the diversity of mechanisms that can lead to cascades growing large and thus focused on relatively successful ones, examining less successful cascades may contribute to a more nuanced understanding of cascade growth.
For example, some modifications of the nomination protocol (e.g., variations in the number of friends one is asked to nominate) were more widespread than others, and may correspond to higher-yield instructions.
More broadly, one could study the evolution of diffusion protocols (e.g., in terms of evolutionary strategies using $r/K$ selection theory \cite{pianka1970r}), both within individual cascades and across classes of cascades.

\section{Acknowledgements}
This work was supported in part by the ARO MURI, Carleton College, a Simons Investigator Award, the Stanford Data Science Initiative, the University of Cambridge, and NSF grant 1741441.

\bibliographystyle{aaai}
\linespread{0.98}\selectfont
\bibliography{refs}

\end{document}